\documentstyle[prb,aps,epsf]{revtex}
\draft

\begin{document}

\title{Restricted and unrestricted Hartree-Fock calculations of
conductance for a  quantum point contact}

\author{O. P. Sushkov}

\address{School of Physics, University of New South Wales,\\
 Sydney 2052, Australia}

\maketitle

\begin{abstract}
Very short quantum wires (quantum contacts) exhibit a conductance 
structure at a value of conductance close to $0.7 \times 2e^2/h$.
It is  believed that the structure arises due to the electron-electron 
interaction, and it is also related to electron spin. 
However details of the mechanism of the structure are not quite clear.
Previously we approached the problem within the restricted Hartree-Fock 
approximation. This calculation demonstrated a structure similar to that 
observed experimentally.
In the present work we perform restricted and unrestricted Hartree-Fock
calculations to analyze the validity of the approximations.
We also consider dependence of the effect on the electron density in leads.
The unrestricted Hartree-Fock method allows us to
analyze trapping of the single electron within the contact.
Such trapping would result in the Kondo model for the ``0.7 structure''.
The present calculation confirms the spin-dependent bound state picture
and does not confirm the Kondo model scenario.
\end{abstract}


The quantized conductance $G=nG_2$, $n=1,2,3,...$, $G_2=2e^2/h$,
through a narrow quantum point contact was
discovered in 1988 \cite{Wharam,Wees}. This quantization can be  
understood within a one-dimensional (1D)
non-interacting electron gas picture, see e.g. Ref.\cite{LB}.
In the present work we are interested in a deviation from the
integer quantization. This deviation, the  so called 
``0.7 structure'' has been found in experimental works \cite{Thomas,Thomas1}.
The structure is a shoulder-like feature or a narrow plateau at
$G\approx 0.7G_2$. More recent work demonstrates that there are some
above barrier excitations related to the structure \cite{Kristensen}, and
that the structure evolves down to $G\approx 0.5G_2$ in longer quantum 
contacts \cite{Reilly}.
Dependence of the structure on the longitudinal magnetic field
has been studied already in the pioneering work \cite{Thomas}. This study 
clearly demonstrated that the effect is somehow related to the electron 
spin. Authors of a recent experimental work \cite{CM} argue that the structure 
signals formation of a Kondo-like correlated spin state.

There have been suggestions to explain the ``0.7 structure'' 
by spontaneous magnetization of the 1D quantum wire
\cite{Chuan,Calmels,Zabala,Spivak,Bychkov}, or by formation of a 
two-electron bound state with nonzero total spin \cite{Flamb,Rejec}.
These suggestions implicitly assume that 2D leads connected to the
contact are qualitatively important
for the effect because there is the rigorous Lieb-Mattis theorem 
\cite{Lieb} that claims that the ground state of a 1D many-body system 
has zero spin.

A Hatree-Fock calculation of the conductance has been performed
in the Ref. \cite{Sushkov}. This calculation demonstrated a structure 
similar to that observed experimentally. The ground state has zero spin
in accordance with the Lieb-Mattis theorem, but nevertheless the structure 
found in the calculation is intrinsically related to the spin because it 
disappears without account of the exchange electron-electron Coulomb 
interaction. The structure is related to the formation of the charge density 
wave within the contact or in other words to the spin-dependent bound state
within the contact. 

The present work  has been stimulated by the recent 
suggestion that the 0.7-structure signals formation of a Kondo-like 
correlated spin state \cite{CM}, see also Ref. \cite{MHW}.
 The restricted Hartree-Fock (RHF) approximation employed in 
Ref. \cite{Sushkov} is not sufficient to follow this suggestion. However the
unrestricted Hartree-Fock (UHF) approach can shed light on the problem.
In the present work we consider only zero temperature case.
The RHF method implies that spin up and spin
down single electron orbitals are identical while in the UHF
method those orbitals are completely independent.
The RHF is explicitly rotationally invariant, but it is not very effective
in accounting for electron-electron correlations. The UHF is much better at
accounting for the correlations, but it violates the rotational invariance.
There is no doubt that even UHF cannot account for the long range Kondo-like
dynamics. However it can indicate localization of a single electron within the
contact. This would immediately imply the Kondo-like dynamics.
Our calculation shows that such localization can take place in longer contacts
and at low electron density in  leads. However it always leads to a very
special dependence of conductance on the gate voltage which is different
from that observed experimentally.
In the regime when the dependence of conductance on the gate voltage is
similar to the experimental one the results of RHF and UHF
are practically identical and this indicates the validity of both 
approximations.
We also study dependence of the ``0.7 structure'' on the electron density 
in the leads. The structure disappears at high density and it is getting more 
pronounced at the low density in a qualitative agreement with experiment.

It is well known, see Ref. \cite{LB}, that in  the independent particle 
approximation, i.e. in the case of an ideal electron gas, 
the conductance for a given transverse channel is proportional
to the barrier transmission probability at Fermi energy $T$,
\begin{equation}
\label{GT}
G={{2e^2}\over{h}} T.
\end{equation}
In case of interacting particles this formula should be also valid because
before and after the potential barrier the density of electrons
is high enough, and hence the interaction is negligible.
However one cannot use a single particle description to 
calculate the transmission probability $T$ because in the vicinity of the 
barrier the electron density is low, and hence the many-body effects are very 
important.
To calculate the transmission probability $T$ the following method is
applied. Consider electrons on a 1D ring of the length $L$ with a potential 
barrier of the length $l$ somewhere on the ring. It is important that
$L \gg l$.
There is no  current in the ground state of the system. Now let us apply a
magnetic flux through the ring. This flux induces the electric current.
Note that it is not a real magnetic field, this is a fictitious gauge
field that generates the current without applying any voltage.
It is especially convenient to take the gauge field that provides
the Bohm-Aharonov phase $\varphi=\pi/2$. We use this choice in our 
calculations. The induced current can be calculated  by solving many-body 
Schroedinger equation. It 
can be an exact solution or an approximate one like RHF or UHF.
It has been demonstrated in Ref.\cite{Sushkov} that to find the barrier 
transmission probability at Fermi energy  one has to 
solve the many-body problem twice: without the barrier and with 
the barrier. 
The ratio of electric currents squared gives the transmission
probability, $T=(J_U/J_0)^2$. This formula is valid without an external magnetic
field. Repeating considerations of Ref.\cite{Sushkov} one can prove that
with the magnetic field, i.e. with the spin splitting, the effective
transmission probability is given by
\begin{equation}
\label{TT}
T={1\over{2}}T_{\uparrow}+{1\over{2}}T_{\downarrow}, \ \ \
T_{\sigma}=\left({{J_{U\sigma}}\over{J_{0\sigma}}}\right)^2,
\end{equation}
where $J_{\uparrow \downarrow}$ is the electric current of electrons
with spin up and spin down correspondingly.
Equation (\ref{TT}) can be also applied for UHF calculations without
an external magnetic field.
The relation $T=(J_U/J_0)^2$ has been applied recently  to study
conductance through a system of strongly correlated spinless fermions \cite{P}.
In this work the many-body problem has been treated {\it exactly} via the 
Density Matrix Renormalization Group algorithm.

The Hamiltonian of the many body system we consider is of the form
\begin{equation}
\label{Hmb}
H=\sum_i\left[{{(p_i-{\cal A})^2}\over{2}}+U(x_i)
\right]+{1\over{2}}\sum_{i,j}V(x_i,x_j),
\end{equation}
where indexes $i$ and $j$ numerate electrons, 
$x_i$ is the periodic coordinate on the ring of length $L$ ($0<x<L$), 
and ${\cal A}=\pi/2L$ is the fictitious gauge field.
The electron-electron Coulomb repulsion is of the form
\begin{equation}
\label{Vij}
V(x,y)={1\over{\sqrt{a_t^2+D^2(x,y)}}},
\end{equation}
where $a_t\approx 2$ is the effective width of the transverse channel, 
see Ref. \cite{Sushkov}, and $D(x,y)$ is the length of the shortest arc  
between the points $x$ and $y$ on the ring.
We use atomic units, so distances are measured in unites of Bohr
radius, $a_B=\epsilon\hbar^2/m e^2$, and energies are measured in units
of $E_{unit}=me^4/\hbar^2\epsilon^2$, where $m$ is the effective electron
mass and $\epsilon$ is the dielectric constant. 
For experimental conditions of works 
\cite{Thomas,Thomas1,Kristensen,Reilly,CM} 
these values are the following: $a_B \approx 10^{-2}\mu m$, 
$E_{unit}\approx 10^{-2}eV$.
To model the gate potential
we use the following formula for the potential barrier 
\begin{equation}
\label{UU}
U(x)={{U_0}\over{e^{(|x|-l/2)/d}+1}}, \ \ \ d=l/10.
\end{equation}
Plots of $U(x)$ for l=8,10,12 are shown  in Fig.1.
\begin{figure}[h]
\vspace{-10pt}
\hspace{-35pt}
\epsfxsize=8.cm
\centering\leavevmode\epsfbox{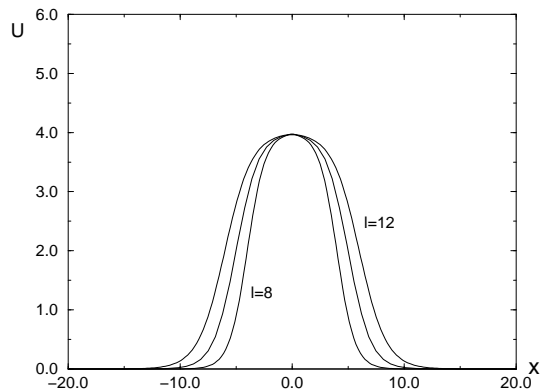}
\vspace{-10pt}
\caption{\it {The ''gate'' potential (\ref{UU}) 
at $U_0=4$ and $l=8,10,12$. }}
\label{Fig1}
\end{figure}
\noindent
To solve the many-body problem described by the Hamiltonian (\ref{Hmb}) we 
use the Hartree-Fock (HF) approximation. In the HF approximation the many 
body wave function is  represented in the form of the Slater determinant of 
single particle orbitals  $\varphi_{i\sigma}(x)$. 
The index $i$ shows the coordinate state of the orbital, and the index
$\sigma = \pm 1/2$ shows the spin state of the orbital.
Each orbital obeys the equation
\begin{equation}
\label{hf}
\hat{h}\varphi_{i\sigma}=\epsilon_{i\sigma}\varphi_{i\sigma},
\end{equation}
where $\epsilon_{i\sigma}$ is the single particle energy and
$\hat{h}$ is the HF Hamiltonian
\begin{eqnarray}
\label{Hhf}
\hat{h}\varphi_{i\sigma}(x)&=&\left({{(p-{\cal A})^2}\over{2}}+U_{eff}(x)
\right)\varphi_{i\sigma}(x)
-\sum_j\int\varphi_{j\sigma}^*(y)\varphi_{i\sigma}(y)V(x,y)dy\varphi_{j\sigma}
(x),\\
U_{eff}&=&U(x)+ \sum_{j\sigma}\int|\varphi_{j\sigma}(y)|^2 V(x,y)dy
.\nonumber
\end{eqnarray}
The summations are performed over all filled orbitals. 
In the Restricted Hartree-Fock (RHF) method an additional constraint,
$\varphi_{i\uparrow}(x)=\varphi_{i\downarrow}(x)$, is imposed on the solutions
of Eqs. (\ref{Hhf}). This provides rotational invariance of the solution.
In the Unrestricted Hartree-Fock method (UHF) the additional constraint is 
omitted. As a result the UHF method provides much better account of
electronic correlations. The price for this is a spontaneous violation of the
rotational invariance.

For computations we use a finite grid.  
In the grid modification of the Hamiltonian (\ref{Hhf}) the kinetic 
energy $(p-{\cal A})^2\varphi$ is replaced by
$\left[2|\varphi(n)|^2-\varphi^*(n+1)e^{iAh}\varphi(n)
-\varphi^*(n)e^{-iAh}\varphi(n+1)\right]/2h^2$. Here $h$ is the spacing of the
grid and $\varphi(n)$ is the wave function on site $n$ of the grid.
The electric current corresponding to the grid Hamiltonian reads
\begin{equation}
\label{jg}
J_{\sigma}=-\sum_j{{i}\over{2h}}\left[\varphi_{j\sigma}^*(n)e^{iAh}
\varphi_{j\sigma}(n+1)-
\varphi_{j\sigma}^*(n+1)e^{-iAh}\varphi_{j\sigma}(n)\right].
\end{equation}
The current is conserved because of the gauge invariance of HF equations.

For computations we use a grid of 400 points on a ring of length 
$L=80$. Total z-projection of the spin is zero, so the number of electrons
with spin up is equal to that with spin down, $N{\uparrow}=N_{\downarrow}$.
We perform calculations for the total number of electrons
$N=N_{\uparrow}+N_{\downarrow}=78, 118, 158$. This corresponds to the following
values of the number density of electrons on the ring: 
$n_0=N/L \approx 1, 1.5, 2$.  This is the effective linear density, therefore 
one cannot compare  $n_0$ quantitatively with the density of electrons in 
real two-dimensional leads used in experiments 
\cite{Thomas,Thomas1,Kristensen,Reilly,CM}. However a qualitative comparison
is possible: the smaller the real density, the smaller $n_0$, and hence the
smaller  the Coulomb screening.
Results of  calculations  for three different values of the barrier length,
$l=8$, $l=10$, $l=12$, and three different values of the electron density in 
the ``leads'', $n_0\approx 1$, $n_0\approx 1.5$, and $n_0\approx 2$, are 
shown in Fig.2.
The transmission probability $T$ is plotted versus the gate potential $U_0$.
Solid lines correspond to the RHF approximation.
\begin{figure}[h]
\vspace{-10pt}
\hspace{-40pt}
\epsfxsize=7.cm
\leavevmode\epsfbox{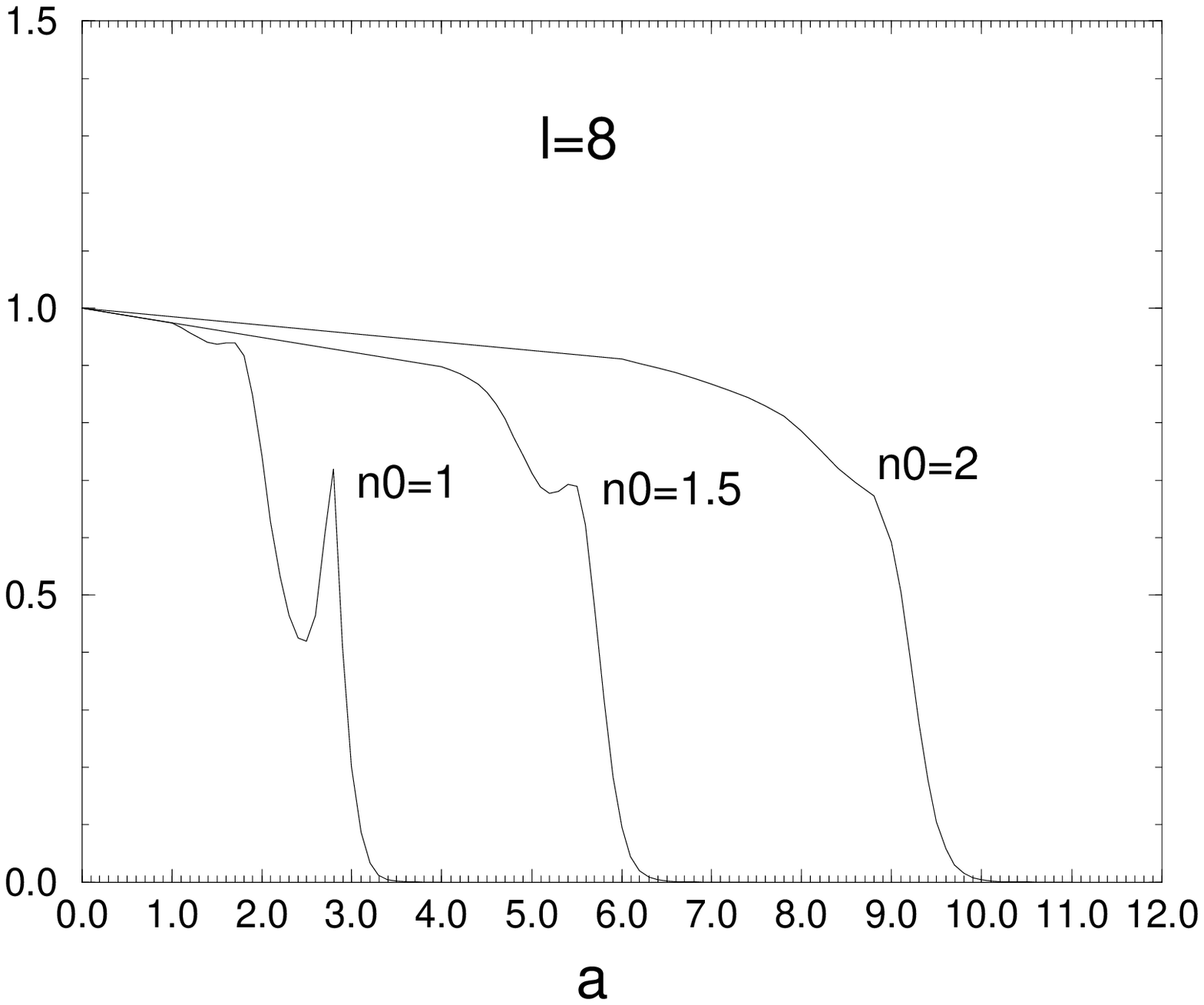}
\hspace{-40pt}
\epsfxsize=7.cm
\leavevmode\epsfbox{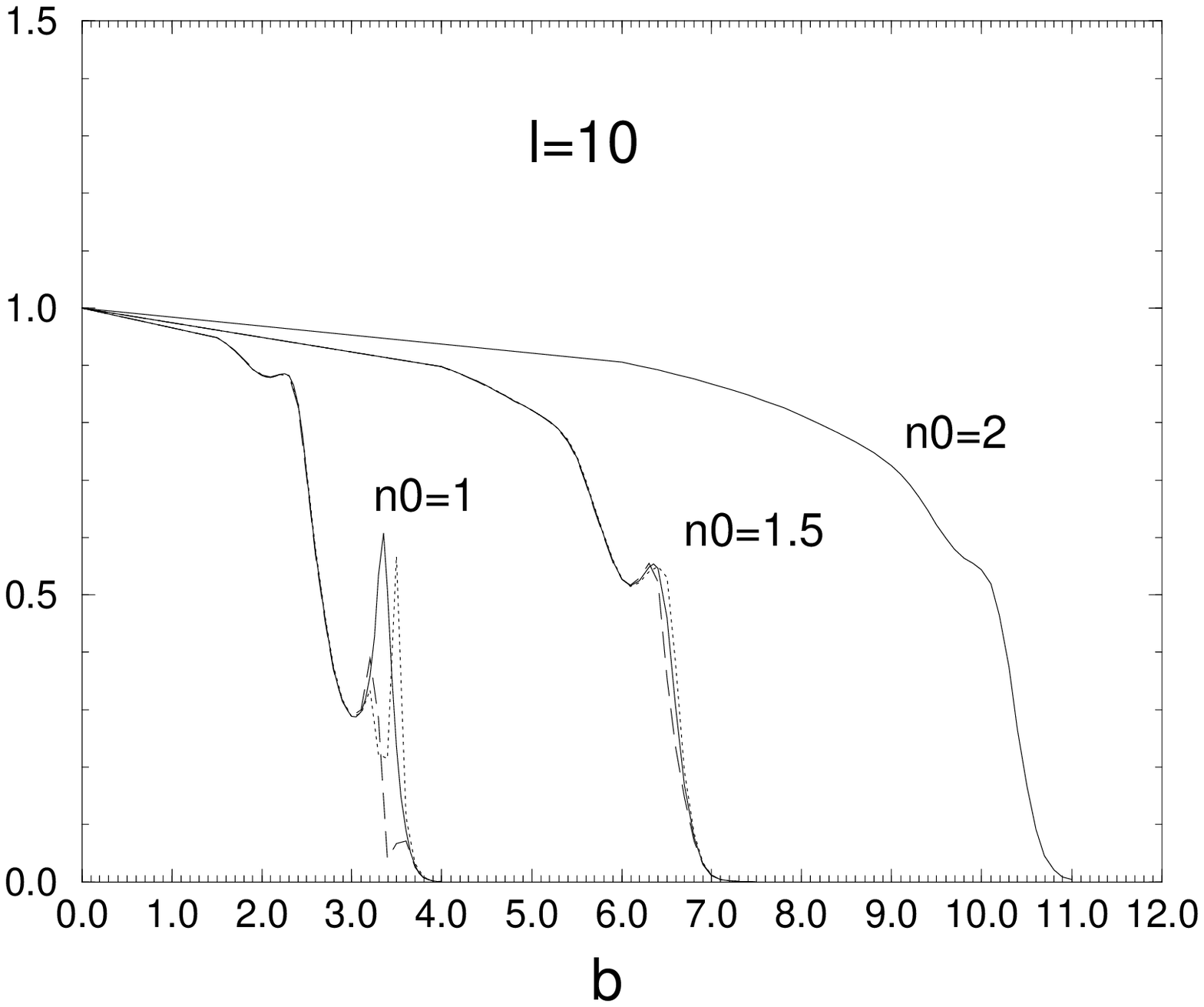}
\hspace{-40pt}
\epsfxsize=7.cm
\leavevmode\epsfbox{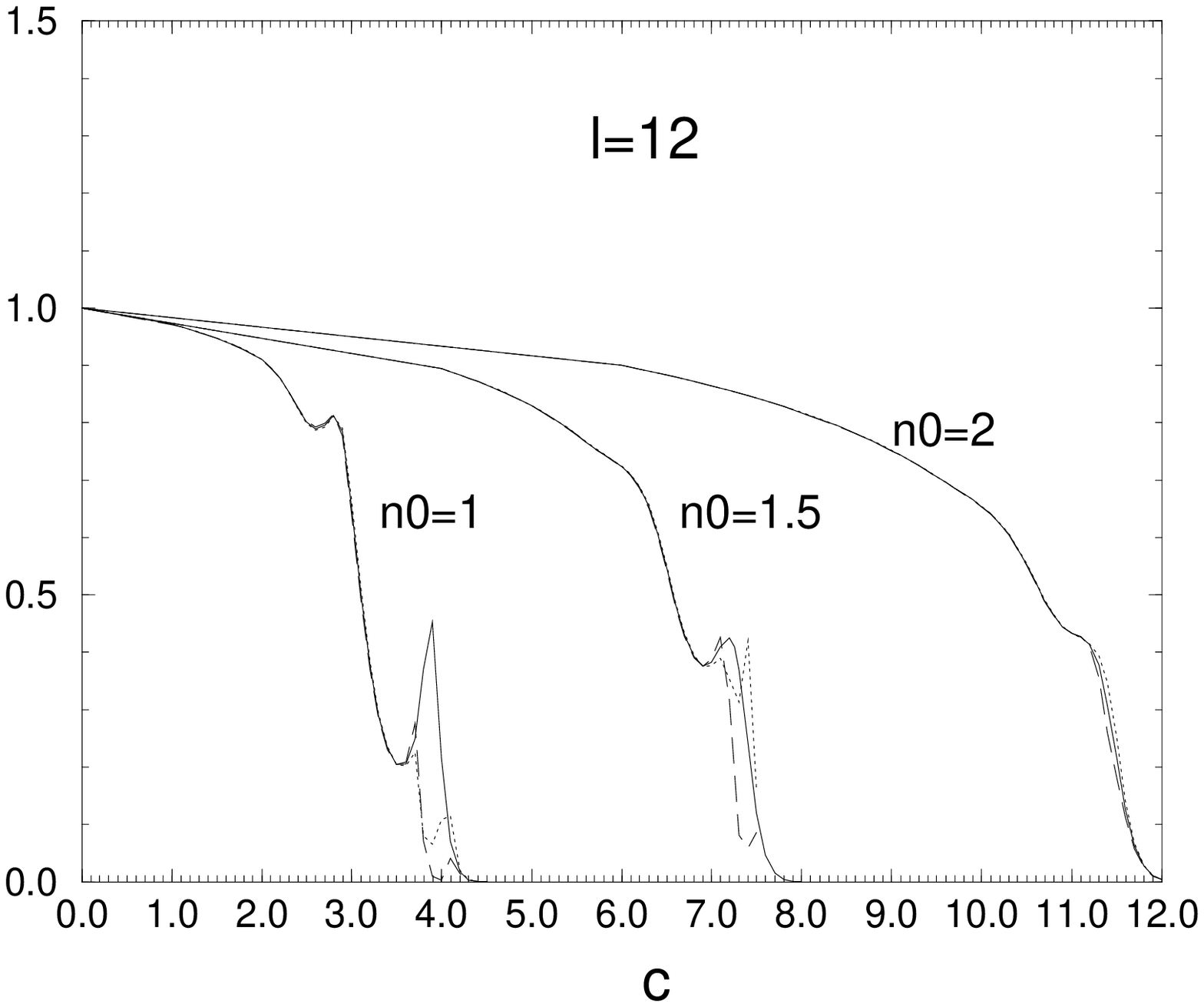}
\vspace{10pt}
\caption{\it {Plots of the transmission probability $T$ versus the gate 
potential
$U_0$ for three different values of the barrier length, $l=8$, $l=10$,
$l=12$, and for three different values of the electron density in the 
``leads'', $n_0=1$, $n_0=1.5$, and $n_0=2$.
Solid lines show results of RHF calculations while dotted and dashed lines
show spin up ($T_{\uparrow}$) and spin down ($T_{\downarrow}$) transmission 
probabilities calculated within the UHF method.
Dotted and dashed lines in Fig.''a'' ($l=8$) are not
distinguishable from solid ones.
}}
\label{Fig2}
\end{figure}
\noindent
The RHF calculation for $n_0\approx 2$ has been performed earlier in Ref.
\cite{Sushkov}.
All the plots presented in Fig.2 clearly demonstrate structures of the
conductance. An important  point is that reduction of the electron density
in leads and hence reduction of screening results in enhancement of the
structure. Another feature is the evolution of the structure down for longer 
contacts.
The results of UHF calculations are shown in the same Fig.1 by
dotted and dashed lines.
The dotted line shows the transmission probability for the spin ``up''
channel and the dashed line shows the same for the spin ``down'' channel.
Certainly the choice of ``up'' and ``down'' is arbitrary, one can
swap the spins. The UHF method always gives two degenerate solutions.
For $l=8$ the UHF results are not presented because they are hardly 
distinguishable from that of the RHF method shown by solid lines.
According to Eq. (\ref{TT}) the observable transmission coefficient 
is the average of $T_{\uparrow}$ and $T_{\downarrow}$.
The results of RHF and UHF methods are very close. To demonstrate
the closeness we also present in Fig.3 plots of electron densities 
$n_{\uparrow}(x)$ and $n_{\downarrow}(x)$ for parameters ($U_0=6.1$, $n_0=1.5$)
and ($U_0=6.4$, $n_0=1.5$) that correspond to two points on the shoulder 
in Fig.2b.
Solid lines represent the RHF density, $n_{\uparrow}(x)=n_{\downarrow}(x)$. 
The dotted line and the dashed line represent UHF densities 
$n_{\uparrow}(x)$ and $n_{\downarrow}(x)$ correspondingly. 
In case (a) ($U_0=6.1$, $n_0=1.5$) the dotted and the dashed lines
are not distinguishable from the solid one. In case (b) ($U_0=6.4$, 
$n_0=1.5$) they are distinguishable, but very close.
\begin{figure}[h]
\vspace{-10pt}
\hspace{40pt}
\epsfxsize=7.cm
\leavevmode\epsfbox{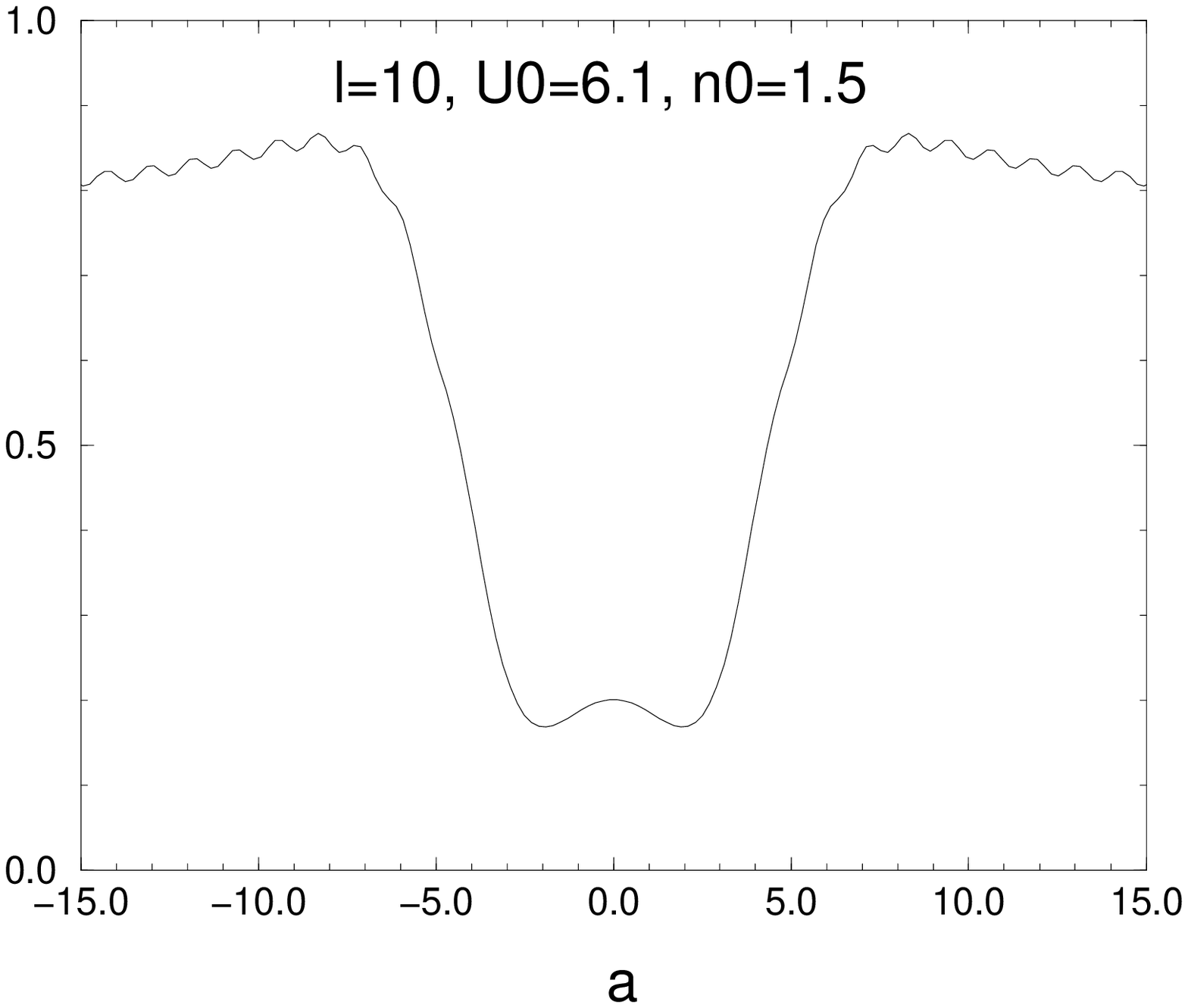}
\hspace{-20pt}
\epsfxsize=7.cm
\leavevmode\epsfbox{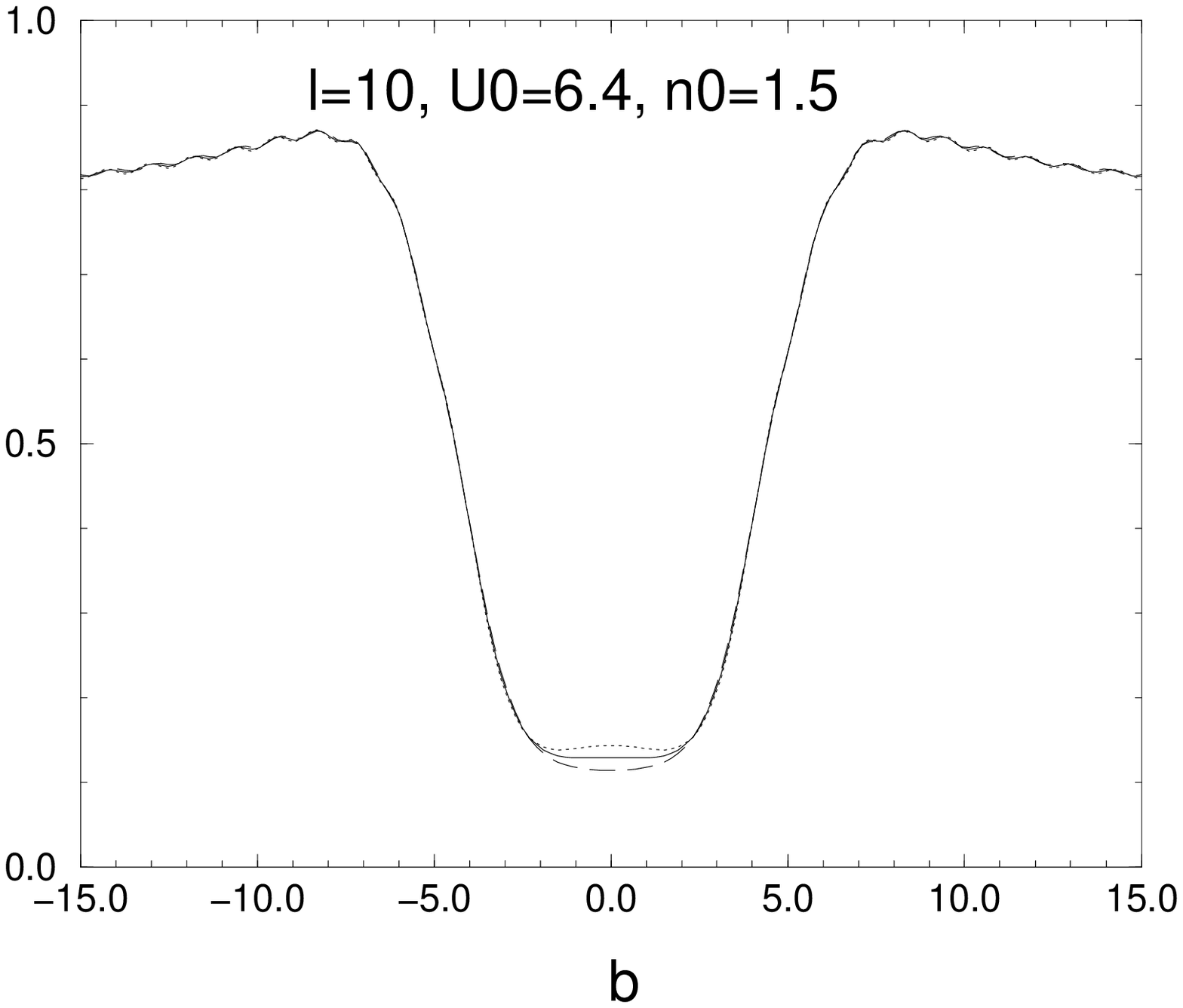}
\hspace{-20pt}
\vspace{5pt}
\caption{\it {Electron densities $n_{\uparrow}(x)$ and $n_{\downarrow}(x)$ 
for parameters ($U_0=6.1$, $n_0=1.5$) and ($U_0=6.4$, $n_0=1.5$) that 
correspond to two points on
the shoulder in Fig.2b. The solid lines show RHF density,
$n_{\uparrow}(x)=n_{\downarrow}(x)$. The dotted line and the dashed line 
show UHF densities $n_{\uparrow}(x)$ and $n_{\downarrow}(x)$ 
correspondingly. In the case (a), ($U_0=6.1$, $n_0=1.5$), the dotted and the 
dashed lines are not distinguishable from the solid one.
}}
\label{Fig3}
\end{figure}
\noindent

According to Fig.2 the RHF and UHF methods  really disagree only at $l=12$, 
$n_0=1$, 
$U_0 > 3.7$: relatively long contact, very low  electron density in leads,
and small conductance. This is the regime where the Kondo model is relevant.
To understand what is going on in this situation we present in Fig.4
plots of electron densities  $n_{\uparrow}(x)$ and $n_{\downarrow}(x)$
at $U_0=4$. As in the above figures, the solid line represents the RHF 
calculation, and  dotted and  dashed lines represent the UHF calculation.
\begin{figure}[h]
\vspace{-10pt}
\hspace{-35pt}
\epsfxsize=7.cm
\centering\leavevmode\epsfbox{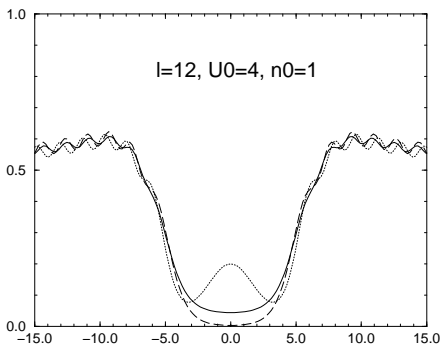}
\vspace{-10pt}
\caption{\it {Electron densities 
for parameters $l=12$, $n_0=1$, and $U_0=4$ that correspond  to the 
structure in the left curve in Fig.2c.
The solid line shows the RHF density, $n_{\uparrow}(x)=n_{\downarrow}(x)$. 
The dotted line and the dashed line show UHF densities $n_{\uparrow}(x)$ 
and $n_{\downarrow}(x)$ correspondingly.
}}
\label{Fig4}
\end{figure}
\noindent
Clearly in this situation the RHF approximation is wrong. According to the
UHF calculation the spin down electron density within the contact is
practically zero and, on the other hand, there is one spin up electron 
localized 
in the contact. There is no doubt that in this case dynamics of the contact 
is Kondo-like. In this case even the UHF method does not give a correct 
transmission  coefficient because the method does not take into account
long-range Kondo dynamics. However, fortunately, the answer is well known; the 
transmission coefficient is peaked up to unity, see Ref. \cite{GR}. So, the 
correct plot of the transmission
coefficient at $l=12$ and $n_0=1$ coincides with that presented in Fig.2c
for $U_0 < 3.7$, and then there is a narrow peak up to $T=1$ at 
$U_0\approx 4$. It is interesting to note that the transmission coefficients
calculated within the Hartree-Fock approximation for $n_0=1$ and for
shorter contacts (Fig.2a,b) have a qualitatively similar dependence:
deep minimum and a narrow peak. This similarity clearly demonstrates how the 
several-electron bound state that can be
assessed by the Hartree-Fock method (Fig.2a, $n_0=1$) evolves to the
multi-electron Kondo bound state that cannot be assessed by this method 
(Fig.2c, $n_0=1$). 

It is interesting that for longer contacts one can trap more
than one electron in the contact. 
To illustrate this in Fig.5 we show UHF electron densities 
$n_{\uparrow}(x)$ and $n_{\downarrow}(x)$ for a contact of length
$l=20$ and density in the leads $n_0=0.56$. In this case the two
electron solution of the type shown in Fig.5a is realized at the gate
potential $1.3 < U_0 < 1.85$, then at $1.85 < U_0 < 2.15$
the solution ``jumps'' to the single electron state shown in Fig.5b. At 
the higher gate potential there are no electrons in the contact.
So adjusting the length of the contact, the density of electrons in leads, and
the gate potential one can pin within the contact a single electron like it 
is shown in Fig.4 and Fig.5b or even the two electron ``molecule'' 
shown in Fig.5a.  However, before getting to this very strongly correlated
regime the transmission probability dips down to the value of few percent.
At most the probability in the dip is 20\% as it is shown 
in Fig.2c ($n_0=1$). 
There are no such dips in experimental data.
Therefore it is unlikely that Kondo dynamics can be 
relevant to the effects observed in
works \cite{Thomas,Thomas1,Kristensen,Reilly,CM}.
On the other hand the plots shown in Figs.2a,b,c for $n_0=1.5$ and $n_0=2$
look very similar to the experimental data.
Structures on these plots are related to the few-electron spin dependent
bound state. The closest physical analogy in this case is probably the Peierls 
spin-density instability.

\begin{figure}[h]
\vspace{-10pt}
\hspace{40pt}
\epsfxsize=7.cm
\leavevmode\epsfbox{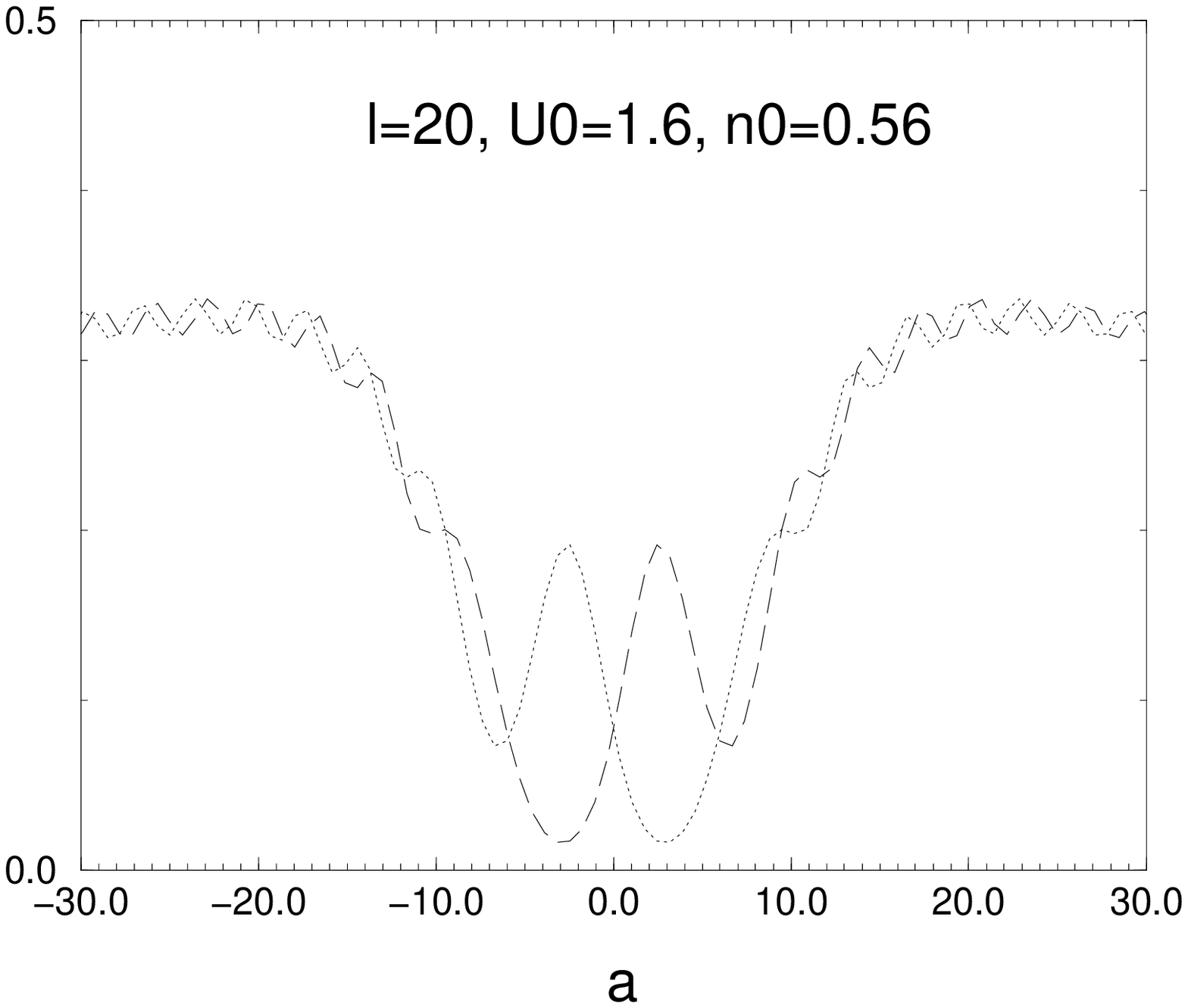}
\hspace{-20pt}
\epsfxsize=7.cm
\leavevmode\epsfbox{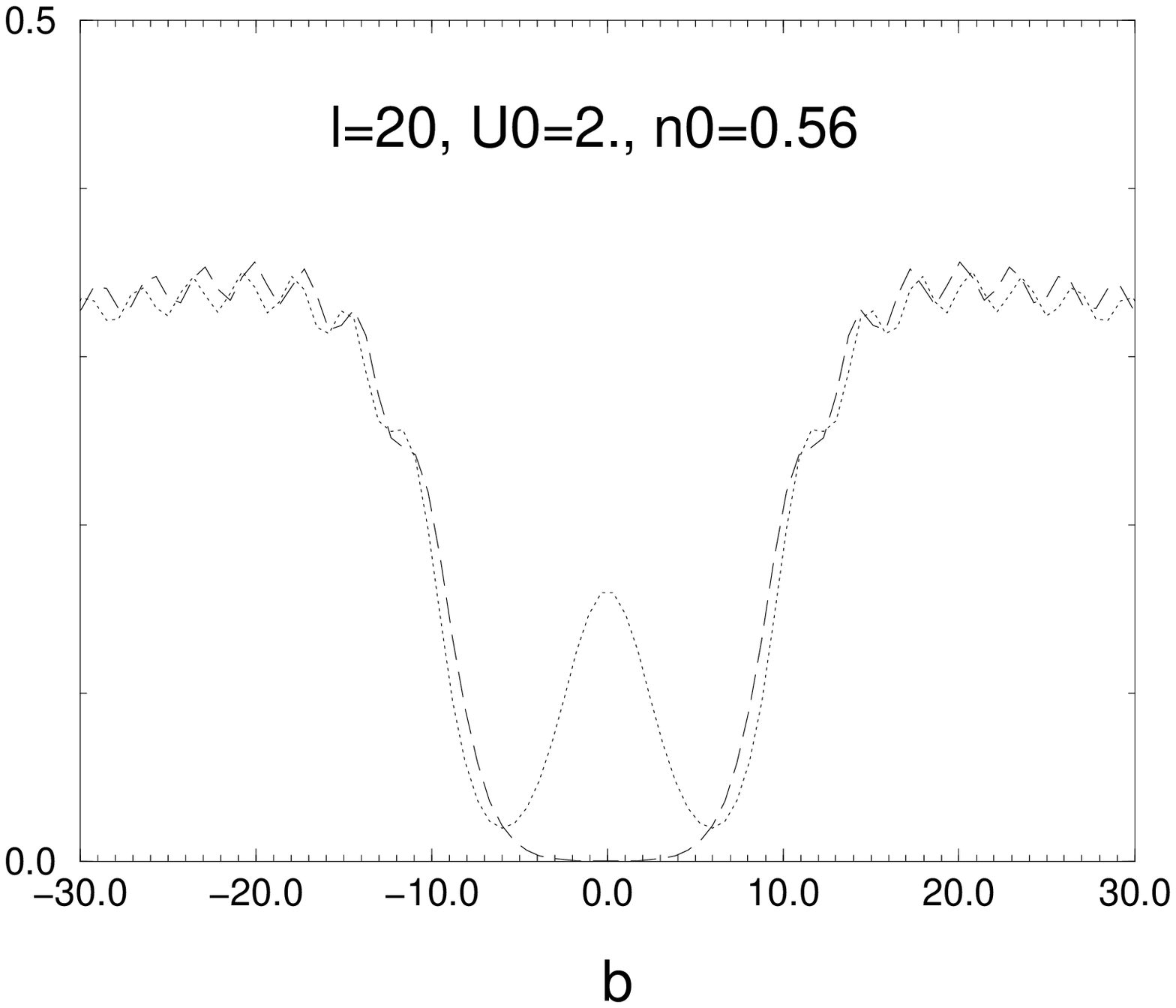}
\hspace{-20pt}
\vspace{10pt}
\caption{\it {UHF electron densities $n_{\uparrow}(x)$ (dotted line)
and $n_{\downarrow}(x)$ (dashed line)
for  a ``very'' long contact, $l=20$, and for a ``very'' low electron density 
in leads, $n_0=0.56$.
}}
\label{Fig5}
\end{figure}

In conclusion, within a one-dimensional model we have analyzed the 
conductance of a short quantum contact at zero temperature.
Restricted (RHF) and unrestricted (UHF) Hartree-Fock  methods have been
used in the analysis. Both methods clearly  demonstrate structures very 
similar to that observed in Refs. \cite{Thomas,Thomas1,Kristensen,Reilly,CM}. 
Agreement 
between RHF and UHF methods confirms the validity of both approximations. 
The conductance  structure is related to the charge density wave developed in 
the contact. This is a spin dependent effect because without the exchange
interaction the structure disappears, so this is a kind of spin-dependent 
bound state within the contact. 
Reduction of the electron density in the leads and hence reduction of the 
screening results in enhancement of the structure. The structure evolves 
down for longer contacts.

Having the contact long enough, the density of electrons in the leads low 
enough,
and adjusting the gate potential one can pin within the contact a single 
electron  or even a two electron ``molecule''. The single electron
would  imply  Kondo-like dynamics as has been suggested in Refs. 
\cite{CM,MHW}. However, according to our calculations, 
before getting to this regime the transmission probability 
as a function of the gate voltage dips down to at least
20\%. Such dip has never been observed experimentally.
Therefore it is unlikely that Kondo dynamics can be 
relevant to the effects observed in
works \cite{Thomas,Thomas1,Kristensen,Reilly,CM}.

I am grateful to  P. G. Silvestrov for very helpful 
discussions. The work has been started at the Institute for Theoretical
Physics at the University of California Santa Barbara, supported by the
National Science Foundation under Grant No PHY99-07949. The work has been
completed at INT at the University of Washington, supported
DOE Grant DE-FG03-00ER41132.  I acknowledge both
institutes for hospitality and support.

\end{document}